%% file: main.tex
\documentclass[12pt]{article}	         

\input{medphy.tex}

\usepackage{amsmath}
\usepackage{booktabs} 
\usepackage{gensymb} 
\usepackage{siunitx} 
\usepackage{caption}
\usepackage{tabularx}
\usepackage{multirow}
\usepackage{makecell}
\usepackage{float}
\usepackage[final]{changes}
\usepackage{listings}
\lstdefinelanguage{json}{
    basicstyle=\ttfamily\scriptsize,
    showstringspaces=false,
    breaklines=true,
    frame=single,
}
\usepackage{xcolor}
\DeclareSIUnit\mGy{mGy}

\begin{document}


\cen{\sf {\Large {\bfseries DoseRAD2026 Challenge dataset: AI accelerated photon and proton dose calculation for radiotherapy}} \\  
\vspace*{10mm}

Fan~Xiao\textsuperscript{1},
Nikolaos~Delopoulos\textsuperscript{1},
Niklas~Wahl\textsuperscript{2,3},
Lennart~Volz\textsuperscript{4},
Lina~Bucher\textsuperscript{5,6,2,3},
Matteo~Maspero\textsuperscript{7},
Miguel~Palacios\textsuperscript{8},
Muheng~Li\textsuperscript{9},
Samir~Schulz\textsuperscript{5,10,2,3},
Viktor~Rogowski\textsuperscript{11},
Ye~Zhang\textsuperscript{9},
Zoltan~Perko\textsuperscript{12,13},
Christopher~Kurz\textsuperscript{1},
George~Dedes\textsuperscript{1,14},
Guillaume~Landry\textsuperscript{1,15,16}, 
Adrian~Thummerer\textsuperscript{1,17,18,a} \\
\vspace{2mm}
\textsuperscript{1} Department of Radiation Oncology, LMU University Hospital, LMU Munich, Munich, Germany \\
\textsuperscript{2} Division of Medical Physics in Radiation Oncology, German Cancer Research Center (DKFZ), Heidelberg, Germany \\
\textsuperscript{3} National Center for Radiation Research in Oncology (NCRO), Heidelberg Institute for Radiation Oncology (HIRO), Heidelberg, Germany \\
\textsuperscript{4} GSI Helmholtzzentrum für Schwerionenforschung, Biophysics Department, Darmstadt, Germany \\
\textsuperscript{5} Department of Physics and Astronomy, Heidelberg University, Heidelberg, Germany \\
\textsuperscript{6} Institute of Biomedical Engineering, Karlsruhe Institute of Technology, Baden-Württemberg, Germany \\
\textsuperscript{7} Department of Radiotherapy, University Medical Center Utrecht, Utrecht, Netherlands \\
\textsuperscript{8} Department of Radiation Oncology, Amsterdam UMC location Vrije Universiteit Amsterdam, Cancer Center Amsterdam, Cancer Treatment and Quality of Life, Amsterdam, Netherlands \\
\textsuperscript{9} Center for Proton Therapy, Paul Scherrer Institute (PSI), Villigen, Switzerland \\
\textsuperscript{10} Department of Radiation Oncology, University Hospital Heidelberg, Heidelberg, Germany \\
\textsuperscript{11} Radiation Physics, Department of Hematology, Oncology, and Radiation Physics, Skåne University Hospital, Lund, Sweden \\
\textsuperscript{12} Delft University of Technology, Department of Radiation Science and Technology, Delft, Netherlands \\
\textsuperscript{13} Radformation Inc., New York, US \\
\textsuperscript{14} Department of Medical Physics, Faculty of Physics, Ludwig-Maximilians-Universit\"at M\"unchen, Munich, Germany \\
\textsuperscript{15} {German Cancer Consortium (DKTK), partner site Munich, a partnership between DKFZ and LMU University Hospital, Germany} \\
\textsuperscript{16} {Bavarian Cancer Research Center (BZKF), Munich, Germany} \\
\textsuperscript{17} Department of Radiation Oncology, Inselspital, Bern University Hospital and University of Bern, Bern, Switzerland \\
\textsuperscript{18} Department of Digital Medicine, Faculty of Medicine, University of Bern, Bern, Switzerland.
\vspace{2mm} \\
\textsuperscript{a} Author to whom correspondence should be addressed: \href{mailto:@unibe.ch}{adrian.thummerer@unibe.ch} \\
\vspace{5mm}
}

\pagenumbering{roman}
\setcounter{page}{1}
\pagestyle{plain}

\clearpage
\begin{abstract}

{\bf Purpose:}
Accurate dose calculation is a fundamental component of modern radiotherapy, enabling precise tumor irradiation while sparing surrounding healthy tissue. With the increasing adoption of \ac{mri}-guided radiotherapy and growing interest in real-time adaptive radiotherapy, there is a demand for fast and accurate dose calculation techniques on both \ac{ct} and \ac{mri}. To facilitate research in this area, the DoseRAD2026 dataset and accompanying challenge introduce a publicly available benchmark dataset comprising paired CT and MRI data with corresponding beam-level photon and proton \ac{mc} dose distributions, enabling the development and systematic evaluation of advanced dose calculation and prediction methods.
\\

{\bf Acquisition and validation methods:}
The DoseRAD2026 dataset comprises paired CT and MRI data from 115 patients (75 training, 40 testing) treated on a hybrid \ac{mri}-linear accelerator (\ac{mri}-linac) for thoracic or abdominal lesions and is derived from the SynthRAD2025 dataset. Image pre-processing included deformable image registration, air-cavity correction and resampling. Ground-truth photon (6 MV) and proton dose distributions were generated using open-source \ac{mc} dose calculation algorithms and resulted in a training dataset of 40,500 photon beams and 81,000 proton beamlets.
\\

{\bf Data format and usage notes:}
The dataset is organized into photon and proton subsets, each comprising paired \ac{ct}-\ac{mri} images, beam-level dose distributions, and JSON files specifying patient-specific beam configurations. Data are provided in compressed MetaImage (.mha) format with standardized naming linking dose files to beam parameters in the JSON files. The dataset is released under a CC BY-NC 4.0 license, with the training set available from April 2026 and the test set withheld until March 2030.
\\

{\bf Potential applications:} 
The DoseRAD2026 dataset enables benchmarking and development of fast dose calculation methods for radiotherapy applications. Potential use cases include rapid beam-level dose estimation for photon and proton therapy, \ac{mri}-based dose calculation in \ac{mri}-only and \ac{mri}-guided radiotherapy workflows, and support for real-time adaptive radiotherapy strategies. 
\\
\end{abstract}



\newpage     

\setlength{\baselineskip}{0.7cm}      
\pagenumbering{arabic}
\setcounter{page}{1}
\pagestyle{fancy}
\acresetall

\section{Introduction}
Radiation therapy is indicated for approximately half of all cancer patients, spanning curative, adjuvant, and palliative contexts \cite{barton2014estimating,lievens2020provision}. Photon therapy, such as \ac{vmat}~\cite{otto2008volumetric}, enables highly conformal dose distributions around the target volume through continuous gantry rotation combined with dynamic \ac{mlc} shaping, while sparing surrounding \acp{oar}. Proton therapy, such as \ac{impt}~\cite{lomax2008intensity}, offers a physically distinct advantage through the Bragg peak~\cite{wilson1946radiological}, whereby the majority of energy is deposited at a finite tissue depth, affording superior sparing of \acp{oar} distal to the tumor. In both modalities, the accuracy of dose calculation is foundational to treatment plan quality~\cite{ahnesjo1999dose}. As workflows increasingly move toward online (immediately prior to radiation delivery)~\cite{lim2017online,paganetti2021adaptive} and ultimately real-time (continuously during radiation delivery) adaptive~\cite{lombardo2024real,keall2025real}, the computational demand placed on dose engines has grown substantially.

\ac{mc} simulation~\cite{rogers2006fifty} on high quality \ac{ct} is broadly accepted as the reference standard for dose calculation, owing to its accurate physics modelling of particle transport. However, general-purpose \ac{mc} implementations suffer from long computation times, with plan-level runtimes spanning hours to tens of hours on a single \ac{cpu}~\cite{paschal2022monte}. \Ac{gpu} acceleration has substantially narrowed this gap for dose calculations, with recent \ac{gpu}-based \ac{mc} engines reporting plan-level runtimes in the range of a few to tens of seconds~\cite{cheng2023development,holmes2024fast,liu2025gpu}. In parallel, \ac{dl} methods have emerged as a complementary strategy, demonstrating sub-second dose prediction and showing early promise at \ac{mc}-comparable accuracy in both proton~\cite{neishabouri2021long,wu2021improving,pastor2022millisecond,neishabouri2025real} and photon~\cite{kontaxis2020deepdose,fan2020data,pastor2023sub,witte2024deep} therapy. Nevertheless, achieving \ac{mc}-level accuracy and the computational speed required for online or real-time plan adaptation---ideally in the sub-second or millisecond time-frame---remains an open challenge.

\Ac{mri} offers superior soft-tissue contrast and eliminates ionizing radiation exposure compared to \ac{ct} \cite{schmidt2015radiotherapy}, making it valuable for target and \ac{oar} delineation as well as motion management in online adaptive radiotherapy. The clinical introduction of hybrid \ac{mr-linac} systems \cite{lagendijk2014magnetic,kluter2019technical} has enabled high-quality soft-tissue imaging prior to and during photon treatment delivery. Meanwhile, prototype systems integrating proton delivery with in-room \ac{mri}, as well as \ac{mri}-only proton workflows \cite{hoffmann2020mr,pham2022magnetic}, have also been proposed. However, since \ac{mri} does not provide electron density or material composition information, dose calculation in both modalities continues to rely on \ac{ct}. Current clinical practice bridges this gap via \ac{sct} generation\cite{bahloul2024advancements,spadea2021sct}, but this intermediate step introduces additional accuracy potential sources of degradation (difference between \ac{sct} and CT), computational latency, and overhead to the dose calculation pipeline, remaining a bottleneck for true real-time plan adaptation. Direct \ac{dl}-based dose prediction from \ac{mri} offers a conceptually cleaner alternative, collapsing the \ac{sct} intermediate into a single end-to-end inference step. Early investigations for both photon \cite{liu2025mr,xiao2025deep} and proton \cite{li2025proof,tian2026magnetic} modalities have demonstrated dosimetric feasibility, yet progress has been constrained by the absence of large-scale paired datasets providing beam-level dose ground truth on both \ac{ct} and \ac{mri}. 

To address this gap, we introduce the DoseRAD2026 dataset, the first publicly available external radiotherapy dose dataset providing paired \ac{ct}s and \ac{mri}s (adapted from the SynthRAD2025 dataset \cite{thummerer2025synthrad2025}) alongside beam-level photon (6 MV) and proton \ac{mc} dose distributions generated using open-source \ac{mc} simulation code. The DoseRAD2026 challenge will utilize this dataset across four tasks: Task~1 (\ac{ct} to photon dose) targets photon segment (the radiation beam shaped by a given MLC aperture) dose calculation on \ac{ct} for \ac{vmat} delivery in thoracic and abdominal cases, where lung and bowel heterogeneities present a challenging test. Task~2  (\ac{mri} to photon dose) extends this to direct photon dose prediction on \ac{mri}, leveraging the paired \ac{ct}-\ac{mri} dataset to map \ac{mc} doses calculated on \ac{ct} onto \ac{mri}. Task~3 (\ac{ct} to proton dose) addresses proton pencil beam dose calculation on \ac{ct} across a range of proton energies in thoracic and abdominal cases. Task~4  (\ac{mri} to proton dose) targets proton dose prediction directly on \ac{mri}, following the same paired data as in Task~2. By releasing beam-level dose labels, DoseRAD2026 is designed to support fast dose calculation both based on CT and also directly on \ac{mri}.

\section{Acquisition and validation methods}

\subsection{Dataset overview}
The DoseRAD2026 dataset comprises paired \ac{ct}s and \ac{mri}s and corresponding photon and proton dose distributions from patients with thoracic and abdominal malignancies. The imaging data were derived from Center B of the SynthRAD2025 dataset \cite{thummerer2025synthrad2025}, a single-institution subset of the public SynthRAD2025 challenge dataset originally released for synthetic CT generation. This subset was selected for the DoseRAD2026 challenge due to its favorable imaging characteristics. In particular, \ac{mri} acquisitions at center B were performed on an \ac{mr-linac} using a standardized protocol, ensuring consistent patient positioning between \ac{ct} and \ac{mri} as well as an axial field of view covering the full patient anatomy. In contrast, \ac{mri}s from other centers exhibited greater variability in acquisition protocols and imaging characteristics, and were sometimes acquired on diagnostic scanners.

For DoseRAD2026, an additional external dataset acquired on a similar \ac{mr-linac} is used for testing, but these data will not be released publicly due to data protection restrictions. This article mainly covers the data from SynthRAD2025 Center B.

\subsection{Image Dataset}

\ac{mri}–\ac{ct} data from Center B of the SynthRAD2025 dataset originally comprised 140 thoracic and 140 abdominal patients, split 91/14/35 for training, validation and test sets, respectively. For the DoseRAD2026 training set, only the currently public SynthRAD2025 training cases were considered, while for testing, only patients from the currently private SynthRAD2025 test set were used, thereby avoiding any impact on the testing phases of both SynthRAD2025 and DoseRAD2026. Because Tasks 2 and 4 focus on direct \ac{mri}-based dose calculation and thus impose strict requirements on \ac{ct}–\ac{mri} alignment, all cases were manually reviewed in three orthogonal planes (axial, coronal, and sagittal) to assess deformable image registration quality (see Section \ref{preprocessing}), and cases with obvious misregistration were excluded to ensure high spatial correspondence between modalities. After screening, a total of 115 patients were selected for DoseRAD2026: 53 thoracic (39 training, 14 testing) and 62 abdominal cases (36 training, 26 testing).

\subsubsection{Pre-processing}\label{preprocessing}

All SynthRAD2025 data had previously undergone several pre-processing steps, including rigid registration, defacing, resampling, and cropping. A detailed description of the SynthRAD2025 pre-processing pipeline is provided by Thummerer et al \cite{thummerer2025synthrad2025}. For DoseRAD2026, four additional pre-processing steps were performed.

First, \ac{ct}s were deformably registered to the corresponding \ac{mri}s using the ConvexAdam deformable image registration framework \cite{siebert2025convexadam}. The ConvexAdam hyperparameters were manually tuned and are listed in the Supplementary Materials \ref{app:convadam_params}.

Second, for abdomen cases, air cavity correction was applied to \ac{ct}s based on \ac{mri} air segmentations. Air cavities in the deformed \ac{ct}s were first identified by thresholding voxels below -200 HU and then filled with the median HU value of all neighboring voxels of the air cavity. Afterwards, air cavities segmented on the \ac{mri}s were transferred to the deformably registered \ac{ct}s and filled with a constant value of -824 HU \cite{godoy2020impact}. To avoid unrealistically sharp edges at the air-tissue interface, the transition between air and soft tissue HUs was slightly blurred in a 1-voxel band around the air cavity boundary. This mimicked the original appearance of air cavities on \ac{ct}s. \ac{mri} air cavity segmentation was performed using an nnU-Net model \cite{isensee2021nnu} trained on a subset of manually annotated cases (20 patients), and all predicted air cavity segmentations were subsequently reviewed and corrected manually.

Third, to reduce file sizes and \ac{mc} photon dose computation times, \ac{ct}s and \ac{mri}s for the photon tasks (Tasks~1 and~2) were resampled to $2\times2\times2\,\mathrm{mm}^3$, while images for the proton tasks (Tasks~3 and~4) were not resampled and remained on the same $1\times1\times3\,\mathrm{mm}^3$ grid as used for SynthRAD2025.

Finally, patient body contours were segmented on both \ac{ct}s and \ac{mri}s using TotalSegmentator \cite{wasserthal2023totalsegmentator}. The intersection of the two masks was used to define a common foreground region, and voxels outside this region were set to -1024 HU in \ac{ct} and 0 in \ac{mri}. 

\subsubsection{Image acquisition parameters}

All center B \ac{mri}s were acquired on a ViewRay MRIdian 0.35 T \ac{mr-linac} using a balanced steady-state free-precession sequence (bSSFP, TrueFISP), while \ac{ct}s were predominantly acquired on a Canon Medical Aquilion/LB CT scanner. Table~\ref{tab:mri_params} and Table~\ref{tab:ct_params} present detailed imaging parameters of \ac{ct}s and \ac{mri}s of the respective datasets. Per patient imaging parameters are provided as metadata to the dataset release.

\begin{table*}[htbp]
\centering
\footnotesize
\setlength{\tabcolsep}{4pt}
\renewcommand{\cellalign}{cc}
\newcommand{\sm}[1]{{\scriptsize\makecell{#1}}}
\caption{\centering MRI acquisition parameters for thorax and abdomen datasets (train/test splits). Voxel spacing is reported as in-plane resolution (isotropic); percentages indicate proportion of scans with each value.}
\label{tab:mri_params}
\resizebox{\textwidth}{!}{%
\begin{tabular}{@{}l cc@{\hspace{3em}}cc@{}}
\toprule
 & \multicolumn{2}{c}{\textbf{Thorax}} & \multicolumn{2}{c}{\textbf{Abdomen}} \\
\cmidrule(lr){2-3} \cmidrule(lr){4-5}
 & \textbf{Train} & \textbf{Test} & \textbf{Train} & \textbf{Test} \\
\midrule
\textbf{Manufacturer} & \multicolumn{4}{c}{ViewRay} \\
\textbf{Model} & \multicolumn{4}{c}{MRIdian} \\
\textbf{Field strength (T)} & \multicolumn{4}{c}{0.35} \\
\textbf{Sequence} & \multicolumn{4}{c}{balanced steady-state free-precession (bSSFP, TrueFISP)} \\
\textbf{Acquisition} & \multicolumn{4}{c}{2D, breath-hold, no contrast} \\
\textbf{Flip angle (\textdegree)} & \multicolumn{4}{c}{60} \\
\textbf{Echo numbers} & \multicolumn{4}{c}{1} \\
\textbf{Number of averages} & \multicolumn{4}{c}{1} \\
\textbf{Slice thickness (mm)} & \multicolumn{4}{c}{3} \\
\midrule
\textbf{TE (ms)} &
  \sm{1.27\,(84\%)\\ 1.43\,(13\%)\\ 1.62\,(3\%)} &
  \sm{1.27\,(64\%)\\ 1.43\,(29\%)\\ 1.62\,(7\%)} &
  \sm{1.27\,(17\%)\\ 1.43\,(53\%)\\ 1.44\,(11\%)\\ 1.62\,(19\%)} &
  \sm{1.43\,(65\%)\\ 1.44\,(15\%)\\ 1.62\,(19\%)} \\
\addlinespace
\textbf{TR (ms)} &
  \sm{3.00\,(84\%)\\ 3.33\,(13\%)\\ 3.83\,(3\%)} &
  \sm{3.00\,(64\%)\\ 3.33\,(29\%)\\ 3.83\,(7\%)} &
  \sm{3.00\,(17\%)\\ 3.33\,(53\%)\\ 3.36\,(11\%)\\ 3.83\,(19\%)} &
  \sm{3.33\,(65\%)\\ 3.36\,(15\%)\\ 3.83\,(19\%)} \\
\addlinespace
\textbf{Phase enc.\ steps} &
  \sm{225\,(32\%)\\ 232\,(53\%)\\ 200\,(13\%)\\ 207\,(3\%)} &
  \sm{225\,(43\%)\\ 232\,(21\%)\\ 200\,(29\%)\\ 207\,(7\%)} &
  \sm{200\,(53\%)\\ 225\,(14\%)\\ 176\,(11\%)\\ 207\,(19\%)\\ 232\,(3\%)} &
  \sm{200\,(65\%)\\ 176\,(15\%)\\ 207\,(19\%)} \\
\addlinespace
\textbf{BW (Hz/px)} &
  \sm{599\,(32\%)\\ 604\,(53\%)\\ 537\,(13\%)\\ 385\,(3\%)} &
  \sm{599\,(43\%)\\ 604\,(21\%)\\ 537\,(29\%)\\ 385\,(7\%)} &
  \sm{537\,(53\%)\\ 599\,(14\%)\\ 534\,(11\%)\\ 385\,(19\%)\\ 604\,(3\%)} &
  \sm{537\,(65\%)\\ 534\,(15\%)\\ 385\,(19\%)} \\
\addlinespace
\textbf{Voxel spacing (mm)} &
  \sm{1.5 $\times$ 1.5\,(97\%)\\ 1.6 $\times$ 1.6\,(3\%)} &
  \sm{1.5 $\times$ 1.5\,(93\%)\\ 1.6 $\times$ 1.6\,(7\%)} &
  \sm{1.5 $\times$ 1.5\,(81\%)\\ 1.6 $\times$ 1.6\,(19\%)} &
  \sm{1.5 $\times$ 1.5\,(80\%)\\ 1.6 $\times$ 1.6\,(19\%)} \\
\addlinespace
\textbf{Acq.\ matrix} &
  \sm{310 $\times$ 360\,(53\%)\\ 300 $\times$ 334\,(32\%)\\ 266 $\times$ 266\,(13\%)\\ 276 $\times$ 276\,(3\%)} &
  \sm{300 $\times$ 334\,(43\%)\\ 310 $\times$ 360\,(21\%)\\ 266 $\times$ 266\,(29\%)\\ 276 $\times$ 276\,(7\%)} &
  \sm{266 $\times$ 266\,(53\%)\\ 276 $\times$ 276\,(19\%)\\ 300 $\times$ 334\,(14\%)\\ 234 $\times$ 234\,(11\%)\\ 310 $\times$ 360\,(3\%)} &
  \sm{266 $\times$ 266\,(65\%)\\ 234 $\times$ 234\,(15\%)\\ 276 $\times$ 276\,(19\%)} \\
\addlinespace
\textbf{Acq.\ time (s)} &
  \sm{25\,(97\%)\\ 17\,(3\%)} &
  \sm{25\,(93\%)\\ 17\,(7\%)} &
  \sm{25\,(69\%)\\ 23\,(11\%)\\ 17\,(19\%)} &
  \sm{25\,(65\%)\\ 23\,(15\%)\\ 17\,(19\%)} \\
\bottomrule
\end{tabular}
}
\end{table*}

\begin{table*}[htbp]
\centering
\footnotesize
\setlength{\tabcolsep}{4pt}
\renewcommand{\cellalign}{cc}
\newcommand{\sm}[1]{{\scriptsize\makecell{#1}}}
\caption{\centering CT acquisition parameters for thorax and abdomen datasets (train/test splits). Voxel spacing is reported as in-plane resolution (isotropic); percentages indicate proportion of scans with each value.}
\label{tab:ct_params}
\resizebox{\textwidth}{!}{%
\begin{tabular}{@{}l cc@{\hspace{3em}}cc@{}}
\toprule
 & \multicolumn{2}{c}{\textbf{Thorax}} & \multicolumn{2}{c}{\textbf{Abdomen}} \\
\cmidrule(lr){2-3} \cmidrule(lr){4-5}
 & \textbf{Train} & \textbf{Test} & \textbf{Train} & \textbf{Test} \\
\midrule
\textbf{Manufacturer} &
  \sm{Canon Medical\\Aquilion/LB\,(97\%)\\ Siemens\\SOMATOM Drive\,(3\%)} &
  \sm{Canon Medical\\Aquilion/LB} &
  \sm{Canon Medical\\Aquilion/LB} &
  \sm{Canon Medical\\Aquilion/LB} \\
\addlinespace
\textbf{kVp} & 120 & 120 & 120 & 120 \\
\textbf{Exposure time (ms)} &
  500 &
  500 &
  \sm{500\,(94\%)\\ 750\,(6\%)} &
  500 \\
\addlinespace
\textbf{Tube current (mA)} &
  40\,--\,305 &
  40\,--\,140 &
  40\,--\,440 &
  40\,--\,190 \\
\addlinespace
\textbf{Rows} & 512 & 512 & 512 & 512 \\
\textbf{Columns} & 512 & 512 & 512 & 512 \\
\textbf{Slice thickness (mm)} &
  3 &
  3 &
  \sm{3\,(94\%)\\ 5\,(6\%)} &
  3 \\
\addlinespace
\textbf{Voxel spacing (mm)} &
  \sm{1.4 $\times$ 1.4\,(68\%)\\ 1.1 $\times$ 1.1\,(29\%)\\ 1.0 $\times$ 1.0\,(3\%)} &
  \sm{1.4 $\times$ 1.4\,(50\%)\\ 1.1 $\times$ 1.1\,(50\%)} &
  \sm{1.4 $\times$ 1.4\,(78\%)\\ 1.1 $\times$ 1.1\,(17\%)\\ 0.8 $\times$ 0.8\,(3\%)\\ 0.6 $\times$ 0.6\,(3\%)} &
  \sm{1.4 $\times$ 1.4\,(92\%)\\ 1.1 $\times$ 1.1\,(8\%)} \\
\addlinespace
\textbf{Data coll.\ diameter (mm)} &
  \sm{700\,(68\%)\\ 550\,(29\%)\\ 500\,(3\%)} &
  \sm{700\,(50\%)\\ 550\,(50\%)} &
  \sm{700\,(78\%)\\ 550\,(17\%)\\ 500\,(3\%)\\ 320\,(3\%)} &
  \sm{700\,(92\%)\\ 550\,(8\%)} \\
\addlinespace
\textbf{Recon.\ diameter (mm)} &
  \sm{700\,(68\%)\\ 550\,(29\%)\\ 500\,(3\%)} &
  \sm{700\,(50\%)\\ 550\,(50\%)} &
  \sm{700\,(78\%)\\ 550\,(17\%)\\ 418\,(3\%)\\ 320\,(3\%)} &
  \sm{700\,(92\%)\\ 550\,(8\%)} \\
\addlinespace
\textbf{Nr.\ of slices} &
  103\,--\,299 &
  117\,--\,154 &
  20\,--\,403 &
  87\,--\,307 \\
\bottomrule
\end{tabular}
}
\end{table*}

\subsubsection{External test dataset}

Additional test data were collected at an external institution, but are intended solely for the testing phase of the DoseRAD2026 challenge and will not be made publicly available. This dataset originally comprised 21 paired thoracic \ac{ct}s and \ac{mri}s. Similar to the public data, \ac{mri}s were acquired on a ViewRay MRIdian 0.35T MR-Linac using the same bSSFP sequence, a slice thickness of 3 $\,\mathrm{mm}$ and a pixel spacing of $1.6 \,\mathrm{mm} \times 1.6 \,\mathrm{mm}$. \ac{ct}s were acquired on a GE Medical Systems Discovery CT590 RT scanner, with a tube voltage of 120 $\,\mathrm{kV}$, a tube current between 167 and 388 $\,\mathrm{mA}$, an exposure between 12 and 49 $\,\mathrm{mAs}$, a pixel spacing between 0.98 and 1.37 $\,\mathrm{mm}$ and a slice thickness of 2.5 $\,\mathrm{mm}$.

These 21 cases underwent the same pre-processing as described above and were manually screened for deformable image registration quality, yielding a filtered external test set of 7 thoracic patients. Since these data were acquired on the same imaging system with the same bSSFP sequence as the training data, they are well suited for use as an external test set for the \ac{mri}-based DoseRAD2026 tasks.

\subsubsection{Image Dataset Split}

Table \ref{tab:doserad2026} summarizes the dataset size and split for ease of reference.
\\

\begin{table}[htbp]
\centering
\caption{\centering Patient distribution in DoseRAD2026 by anatomical region and dataset split. Training and Testing are from SynthRAD2025 Center B.}
\begin{tabular}{lcccc}
\hline
\textbf{Region} & \textbf{Training} & \textbf{Testing} & \textbf{External Testing} & \textbf{Total} \\
\hline
Thoracic  & 39 & 14 & 7  & 60 \\
Abdominal & 36 & 26 & -- & 62 \\
\hline
\textbf{Total} & \textbf{75} & \textbf{40} & \textbf{7} & \textbf{122} \\
\hline
\\
\end{tabular}

\label{tab:doserad2026}
\end{table}

\subsection{Dose Dataset}

After the pre-processing and manual review, the corrected \ac{ct}s were subsequently converted into mass density (details see Supplementary Materials \ref{app:beam_params}) and elemental composition maps using a scanner-specific calibration curve \cite{schmid2015monte}. All \ac{mc} simulations were implemented using the Geant4 toolkit (version 11.00-patch-03) with the QGSP\_BIC\_EMV physics list, scoring dose-to-medium of each voxel in units of Gray (Gy). Task-specific simulation details for photon and proton beams are described in Sections II.C.1 and 2.

\subsubsection{Photon MC simulation}

\ac{vmat} plans were generated for all 75 training and 40 test patients using a \ac{vmat} implementation \cite{christiansen2018continuous} in the open-source research treatment planning toolkit matRad~\cite{wieser2017development,abbani_2026_19227986} (v3.2.0).  Each plan was optimized based on the patient's \ac{ct} and clinically contoured structure set, using a simplified model configured for an Elekta Versa HD Linac with an Agility 160-leaf \ac{mlc}, in which jaw positions, leaf transmission, and tongue-and-groove effects were not modelled. Optimization followed a three-stage sequential pipeline: fluence map optimization, leaf sequencing via the Siochi algorithm \cite{siochi2007variable} and direct aperture optimization. 

The resulting plan comprised a single coplanar full arc (gantry \ang{-180} to \ang{180}, couch \ang{0}), yielding 181 control points from matRad, of which 180 were retained after removing the duplicate at \ang{180} (coincident with \ang{-180}). Each control point was defined by a gantry angle, \ac{mu}, and \ac{mlc} leaf positions. 180 beam segments were then derived based on the \ac{mlc} aperture of the corresponding control point. Each aperture was discretized into a binary mask on a uniform pixel grid ($1~\text{mm}\times1~\text{mm}$), where a pixel was assigned a value of one if fully open and zero if blocked. For the 75 training patients, this pool of segments was augmented by randomly shifting each \ac{mlc} leaf by $\pm$5\,mm along its direction of travel two times, expanding the pool from 180 to 540 segments per patient (40,500 in total). \Ac{vmat} plan segments from the 40 internal testing cases and 7 external testing cases were extracted without augmentation.

After segment extraction, Geant4 dose simulations with a 6\,MV photon energy spectrum derived from the ELEKTA\_PRECISE\_6MV phase space files provided by the International Atomic Energy Agency \cite{capote2006phase} were implemented (photon energy distribution in Supplementary Materials \ref{app:beam_params}). The virtual source point-to-isocenter distance was set to 100\,cm. At the isocenter plane normal to the beam, a fully open segment mask would correspond to a $40\,\mathrm{cm} \times 40\,\mathrm{cm}$ field, and the segment mask was defined a corresponding virtual plane of this dimension located at the opposite of the virtual source point. This served as a fluence source from which particles were emitted with focus on the virtual source point. The dose grid matched the patient \ac{ct} spacing ($2~\mathrm{mm} \times 2~\mathrm{mm} \times 2~\mathrm{mm}$). A uniform 0.35\,T magnetic field was simulated in Geant4, aligned parallel to the superior-inferior patient direction and confined within a cylindrical region with a 50\,cm diameter \cite{kluter2019technical}.

\begin{figure}[H]
    \centering
    \makebox[\textwidth][c]{\includegraphics[width=\textwidth]{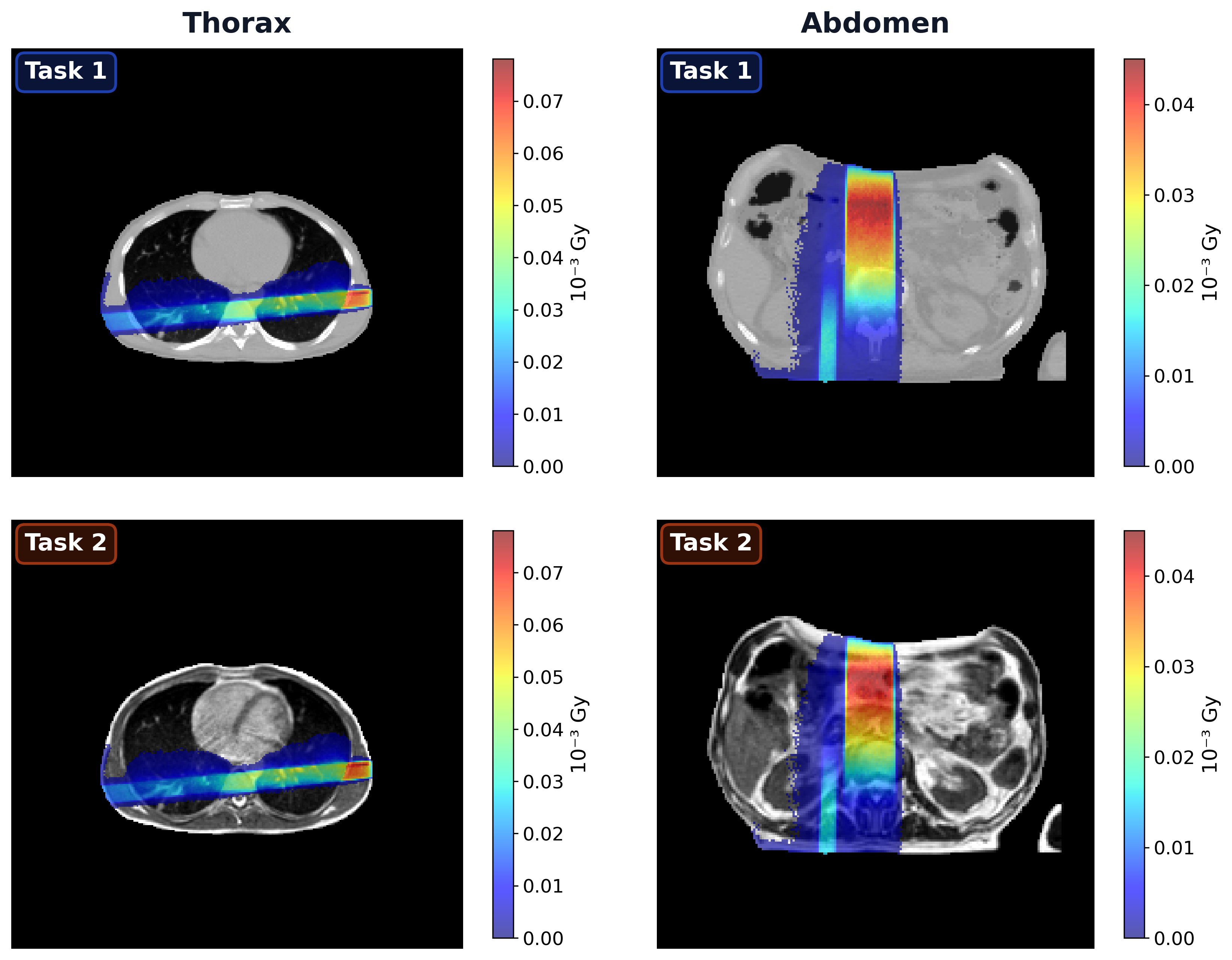}}
    \caption{\centering Representative photon segment dose distributions overlaid on CT (Task~1, top row) and MRI (Task~2, bottom row) images for thoracic (left column) and abdominal (right column) cases.}
    \label{fig:dose_photon_tasks}
\end{figure}

For each training patient, the isocenter from the matRad \ac{vmat} plan was shifted along the superior-inferior axis by $-2$, $0$, and $+2$\,cm, defining three arcs. For each gantry angle, the three variants in the 540-segment pool (1 original + 2 \ac{mlc}-perturbed) were assigned to the three arcs, so that each arc received one unique aperture per angle (180 segments per arc). Segment doses were then simulated at gantry angles every $2^\circ$ for all three arcs, yielding $75~\text{patients} \times 3~\text{arcs} \times 180~\text{angles} = 40{,}500$ training samples in total. Representative photon segment dose distributions in the training dataset for both Task~1 (CT) and Task~2 (MRI) are illustrated in Fig.~\ref{fig:dose_photon_tasks}.

For the testing patients, segment doses were simulated according to each testing plan's control point sequence. Plan doses were then reconstructed by \ac{mu}-weighted accumulation of segment doses. The dose prescriptions to the \ac{ptv} were 70 Gy in 35 fractions for lung patients and 60 Gy in 30 fractions for abdomen patients (2 Gy per fraction in both cases). All \ac{mc} testing plan doses were scaled so that $D_{95\%}(\text{\ac{ptv}}) = 0.95 \times D_\text{prescribed}$. For both training and test sets, each photon segment dose was simulated at a fixed incident fluence of $1\times10^{7}$\,photons/cm$^{2}$, such that the number of simulated photons varies with segment size. 

\vspace{1em}

\subsubsection{Proton MC simulation}

After the HU-to-mass density and elemental composition conversion, stopping powers were computed internally by Geant4 from the assigned material properties using the Bethe--Bloch formalism, with \ac{rsp} (relative to water) optionally extracted in our Geant4 software  (which may be useful for \ac{dl} model training \cite{radonic2024proton}). The initial proton energies ranged from 31.73 to 200.80\,MeV (85 energy levels). A simplified proton beam model was employed to define the energy-dependent beamlet energy spread (sigma  from 0.78 to 6.11\,MeV) and spot size (sigma from 4.00 to 7.99\,mm), under a single-Gaussian lateral approximation. Proton parameters were derived from matRad's native "Generic" proton machine and are listed in Table 4. The proton dose grid matched the patient \ac{ct} spacing ($1 \times 1 \times 3$\,mm) and no magnetic field was simulated.

\begin{figure}[H]
    \centering
    \makebox[\textwidth][c]{\includegraphics[width=\textwidth]{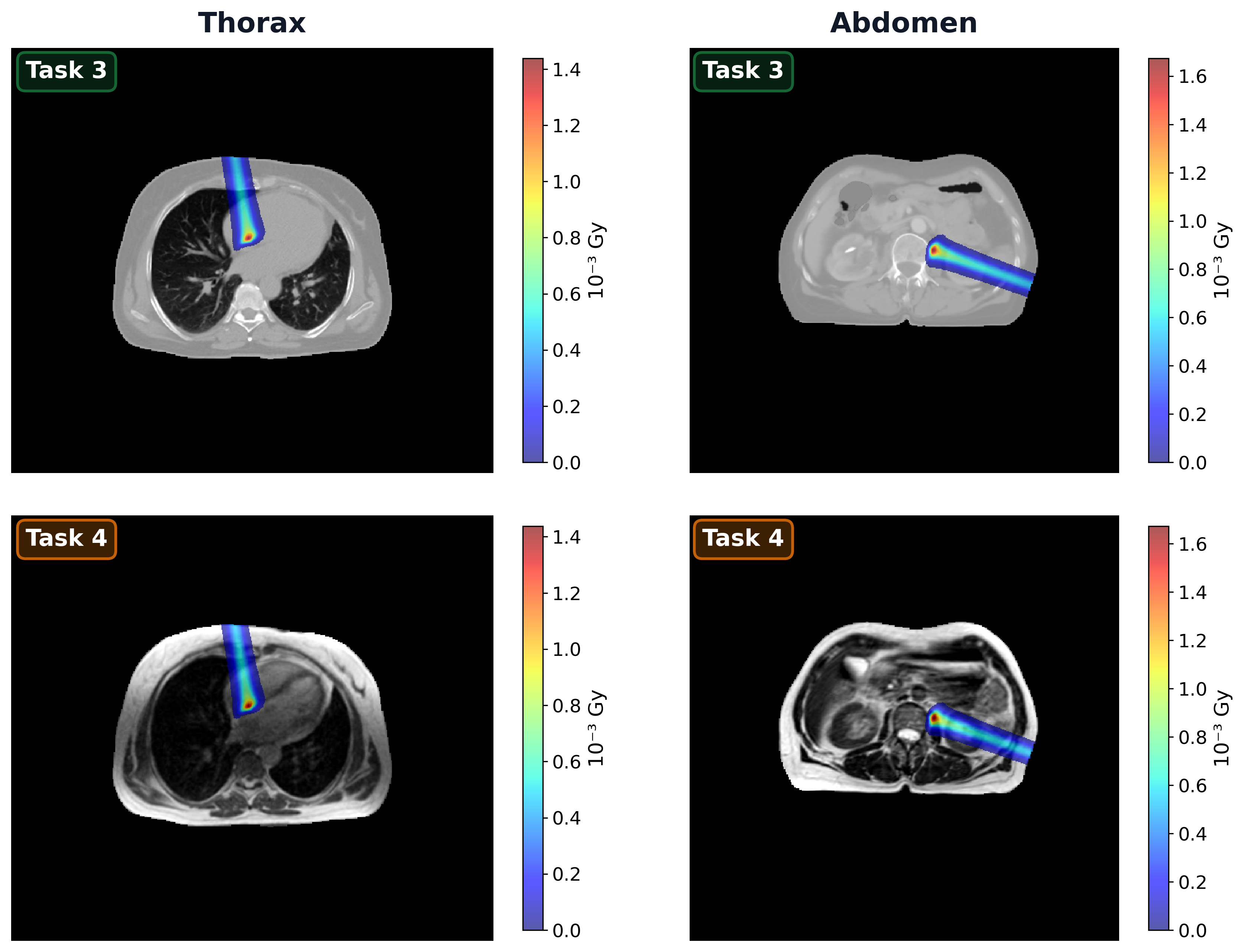}}
    \caption{\centering Representative proton beamlet dose distributions overlaid on CT (Task~3, top row) and MRI (Task~4, bottom row) images for thoracic (left column) and abdominal (right column) cases.}
    \label{fig:dose_proton_tasks}
\end{figure}

For the 75 training patients, a total of 36 gantry angles were simulated in $10^\circ$ increments from $0^\circ$ to $350^\circ$ surrounding the isocenter. Along each gantry direction, proton spots were initialized as parallel pencil beams on a source plane located 100\,cm upstream of the isocenter, approximating the extended virtual source-to-axis distances arising from the scanning magnet positions in clinical proton systems. At each gantry angle, 15 rays were arranged on a $5 \times 3$ \ac{bev} grid, with 5 fixed positions of $\{-40, -20, 0, +20, +40\}$\,mm along the superior-inferior axis, and 3 positions along the lateral axis consisting of $0$\,mm and two randomly sampled offsets from $\{\pm 5k \mid k \in [1,6]\}$\,mm. Each ray was assigned 2 beamlets of distinct energies sampled from Table 4, yielding $15 \times 2 = 30$ beamlets per beam. Across all 75 training patients, this setup produced $75~\text{patients} \times 36~\text{angles} \times 30~\text{beamlets} = 81{,}000$ beamlet dose distributions as the training dose dataset. Representative proton beamlet dose distributions in the training dataset for both Task~3 (\ac{ct}) and Task~4 (\ac{mri}) are illustrated in Fig.~\ref{fig:dose_proton_tasks}.

For the 40 testing patients, 40 \ac{impt} plans were optimized based on the patient's \ac{ct} and clinically contoured structure set using matRad. Plan doses were then calculated as a spot weight-scaled superposition of beamlet doses using Geant4. The \ac{ptv} dose prescriptions were 70 Gy in 35 fractions for lung patients and 60 Gy in 30 fractions for abdomen patients (2 Gy per fraction in both cases). All \ac{mc} testing plan doses were scaled so that $D_{95\%}(\text{\ac{ptv}}) = 0.95 \times D_\text{prescribed}$. For both training and test proton dose simulations, each beamlet dose was simulated with $1\times10^{6}$ primary protons.

\vspace{2em}
\begin{table}[H]
\centering
\caption{\centering Proton beam model parameters: energy, energy spread ($\sigma_E$), and spot size ($\sigma_s$) for 85 energy levels (values rounded to 2 decimal places).}
\label{tab:proton_energies}
\scriptsize
\setlength{\tabcolsep}{5pt}
\begin{tabular}{ccc|ccc|ccc}
\toprule
\textbf{E (MeV)} & $\boldsymbol{\sigma_E}$ \textbf{(MeV)} & $\boldsymbol{\sigma_s}$ \textbf{(mm)} &
\textbf{E (MeV)} & $\boldsymbol{\sigma_E}$ \textbf{(MeV)} & $\boldsymbol{\sigma_s}$ \textbf{(mm)} &
\textbf{E (MeV)} & $\boldsymbol{\sigma_E}$ \textbf{(MeV)} & $\boldsymbol{\sigma_s}$ \textbf{(mm)} \\
\midrule
31.73 & 6.11 & 7.49 & 114.54 & 1.47 & 5.85 & 162.93 & 0.99 & 4.49 \\
36.80 & 5.19 & 7.62 & 116.53 & 1.44 & 5.78 & 164.45 & 0.98 & 4.46 \\
41.38 & 4.56 & 7.70 & 118.49 & 1.42 & 5.70 & 165.96 & 0.97 & 4.44 \\
45.60 & 4.09 & 7.77 & 120.43 & 1.39 & 5.63 & 167.46 & 0.96 & 4.41 \\
49.54 & 3.74 & 7.82 & 122.34 & 1.37 & 5.56 & 168.95 & 0.95 & 4.39 \\
53.25 & 3.45 & 7.86 & 124.23 & 1.34 & 5.50 & 170.43 & 0.94 & 4.36 \\
56.77 & 3.21 & 7.90 & 126.10 & 1.32 & 5.43 & 171.90 & 0.93 & 4.34 \\
60.13 & 3.01 & 7.93 & 127.95 & 1.30 & 5.37 & 173.36 & 0.92 & 4.32 \\
63.35 & 2.84 & 7.95 & 129.78 & 1.28 & 5.31 & 174.81 & 0.91 & 4.30 \\
66.45 & 2.70 & 7.98 & 131.59 & 1.26 & 5.25 & 176.25 & 0.91 & 4.28 \\
69.44 & 2.57 & 7.99 & 133.37 & 1.24 & 5.20 & 177.68 & 0.90 & 4.26 \\
72.33 & 2.46 & 7.99 & 135.15 & 1.22 & 5.15 & 179.10 & 0.89 & 4.24 \\
75.14 & 2.35 & 7.71 & 136.90 & 1.21 & 5.10 & 180.52 & 0.88 & 4.22 \\
77.87 & 2.26 & 7.56 & 138.64 & 1.19 & 5.05 & 181.92 & 0.87 & 4.20 \\
80.53 & 2.18 & 7.42 & 140.35 & 1.17 & 5.00 & 183.32 & 0.87 & 4.18 \\
83.13 & 2.10 & 7.29 & 142.06 & 1.16 & 4.95 & 184.71 & 0.86 & 4.17 \\
85.66 & 2.04 & 7.16 & 143.75 & 1.14 & 4.91 & 186.09 & 0.85 & 4.15 \\
88.13 & 1.97 & 7.03 & 145.42 & 1.13 & 4.87 & 187.46 & 0.84 & 4.14 \\
90.56 & 1.91 & 6.91 & 147.08 & 1.11 & 4.82 & 188.83 & 0.84 & 4.12 \\
92.93 & 1.86 & 6.80 & 148.72 & 1.10 & 4.79 & 190.19 & 0.83 & 4.11 \\
95.26 & 1.81 & 6.69 & 150.35 & 1.08 & 4.75 & 191.54 & 0.82 & 4.09 \\
97.55 & 1.76 & 6.58 & 151.97 & 1.07 & 4.71 & 192.88 & 0.82 & 4.08 \\
99.79 & 1.72 & 6.48 & 153.57 & 1.06 & 4.68 & 194.22 & 0.81 & 4.06 \\
102.00 & 1.68 & 6.38 & 155.16 & 1.05 & 4.64 & 195.55 & 0.80 & 4.05 \\
104.17 & 1.64 & 6.28 & 156.74 & 1.03 & 4.61 & 196.87 & 0.80 & 4.04 \\
106.30 & 1.60 & 6.19 & 158.31 & 1.02 & 4.58 & 198.19 & 0.79 & 4.02 \\
108.41 & 1.57 & 6.10 & 159.86 & 1.01 & 4.55 & 199.49 & 0.79 & 4.01 \\
110.48 & 1.53 & 6.02 & 161.40 & 1.00 & 4.52 & 200.80 & 0.78 & 4.00 \\
112.52 & 1.50 & 5.93 & & & & & & \\
\bottomrule
\end{tabular}
\end{table}

\subsubsection{Post-processing}

All simulated photon and proton dose distributions were masked to zero outside the body contour derived from the deformed \ac{ct}.

\subsubsection{Computational Resources}

The average Geant4 simulation time for a single photon segment dose was approximately 20--30 hours on a single \ac{cpu} core, while a single proton beamlet dose required approximately 2--3 hours. All simulations were carried out on the computing clusters of the GSI Helmholtzzentrum für Schwerionenforschung, the Ludwig Maximilian University (LMU) Hospital Munich and Leibniz-Rechenzentrum (LRZ) utilizing approximately 2{,}000 \ac{cpu} cores in total.

\subsection{Data validation}

\subsubsection{Image dataset validation}

The DoseRAD2026 dataset was derived from the SynthRAD2025 dataset, which already underwent extensive quality control by the initial data providers through visual inspection to ensure completeness and overall data integrity. However, whereas the SynthRAD2025 dataset was designed to represent a heterogenous, multi-center and multi-anatomy cohort, containing a broad range of scanner models and acquisition protocols, the aims of the DoseRAD2026 challenge required special attention to image alignment between \ac{ct} and \ac{mri}. This resulted in the inclusion of only a single center with data from two anatomical regions (thorax and abdomen). All included cases were manually screened and rigorously evaluated for high-quality alignment between \ac{ct}s and \ac{mri}s. This assessment was performed by two medical imaging experts, who visually inspected deformably registered image pairs with particular attention to alignment in bony structures, soft tissue interfaces, and external body contours. For these cases, detailed imaging parameters were systematically extracted from the SynthRAD2025 metadata and are provided as metadata to the DoseRAD2026 dataset to facilitate reproducibility and transparency.

The data inspection further revealed the presence of metal artifacts, including dental implants, in 17 out of 115 patients (13 in the training set and 4 in the test set). These cases were intentionally retained, as they provide a valuable opportunity to assess the robustness and sensitivity of different dose calculation approaches in the presence of clinically relevant imaging artifacts. A detailed list of the 13 patients with metal implants from the training set is provided in the Supplementary Materials \ref{app:metal_ids}.

\subsubsection{Dose dataset validation}

To assess \ac{mc} statistical uncertainty, 10 independent simulations with different random seeds were performed on one representative thoracic and one representative abdominal case, for both photon segment doses and proton beamlet doses. The statistical uncertainty was quantified as the voxel-wise standard deviation across the 10 repeated simulations, evaluated within dose regions exceeding 10\% of $D_{\max}$ to exclude low-dose voxels where relative uncertainty is inherently inflated. For photon segment doses, a primary fluence of $1\times10^{7}$\,photons/cm$^{2}$ yielded an average statistical uncertainty below 3\% in both cases. For proton beamlet doses, simulating $1\times10^{6}$ primary protons resulted in an average statistical uncertainty below 2\%.

All photon dose distributions were verified using \ac{bev} extraction based on the beam direction to match the corresponding dose and visual inspection of the segment shape, whereas all proton dose distributions were checked via \ac{bev} extraction based on the beamlet direction.

\subsection{Code availability}
DoseRAD2026 preprocessing, MC simulation and matRad planning scripts are publicly available and listed in table \ref{tab:code-availability}. 

In addition, analytical pencil beam dose calculation scripts for both photon and proton modalities, built upon matRad's Python variant pyRadPlan~\cite{becher_2026_19489368}, are provided (link see Table \ref{tab:code-availability}). These scripts serve as templates for DoseRAD2026 challenge participants.

\vspace{2em}
\begin{table}[H]
\centering
\caption{\centering Availability of DoseRAD2026 code and scripts.}
\label{tab:code-availability}
\scriptsize
\setlength{\tabcolsep}{5pt}
\begin{tabular}{p{0.30\linewidth} p{0.70\linewidth}}
\toprule
\textbf{Component} & \textbf{Repository URL} \\
\midrule
Pre-processing scripts & \url{https://github.com/DoseRAD2026/preprocessing} \\
Geant4 simulation scripts & \url{https://github.com/DoseRAD2026/geant4-dose-sim} \\
Geant4 toolkit version & \url{https://geant4.web.cern.ch/download/11.0.3.html} \\
matRad VMAT planning scripts & \url{https://github.com/DoseRAD2026/matrad-vmat-example} \\
matRad commit used in this study & \url{https://github.com/e0404/matRad/tree/71c95e96a1ecc7c7e025b9d75240ea009bdbc063} \\
pyRadPlan pencil beam baseline & \url{https://github.com/DoseRAD2026/pyradplan-pb-baseline} \\
\bottomrule
\end{tabular}
\end{table}

\section{Data format and usage notes}

\subsection{Data structure and file formats}

\begin{figure}[H]
    \centering
    \includegraphics[width=\textwidth]{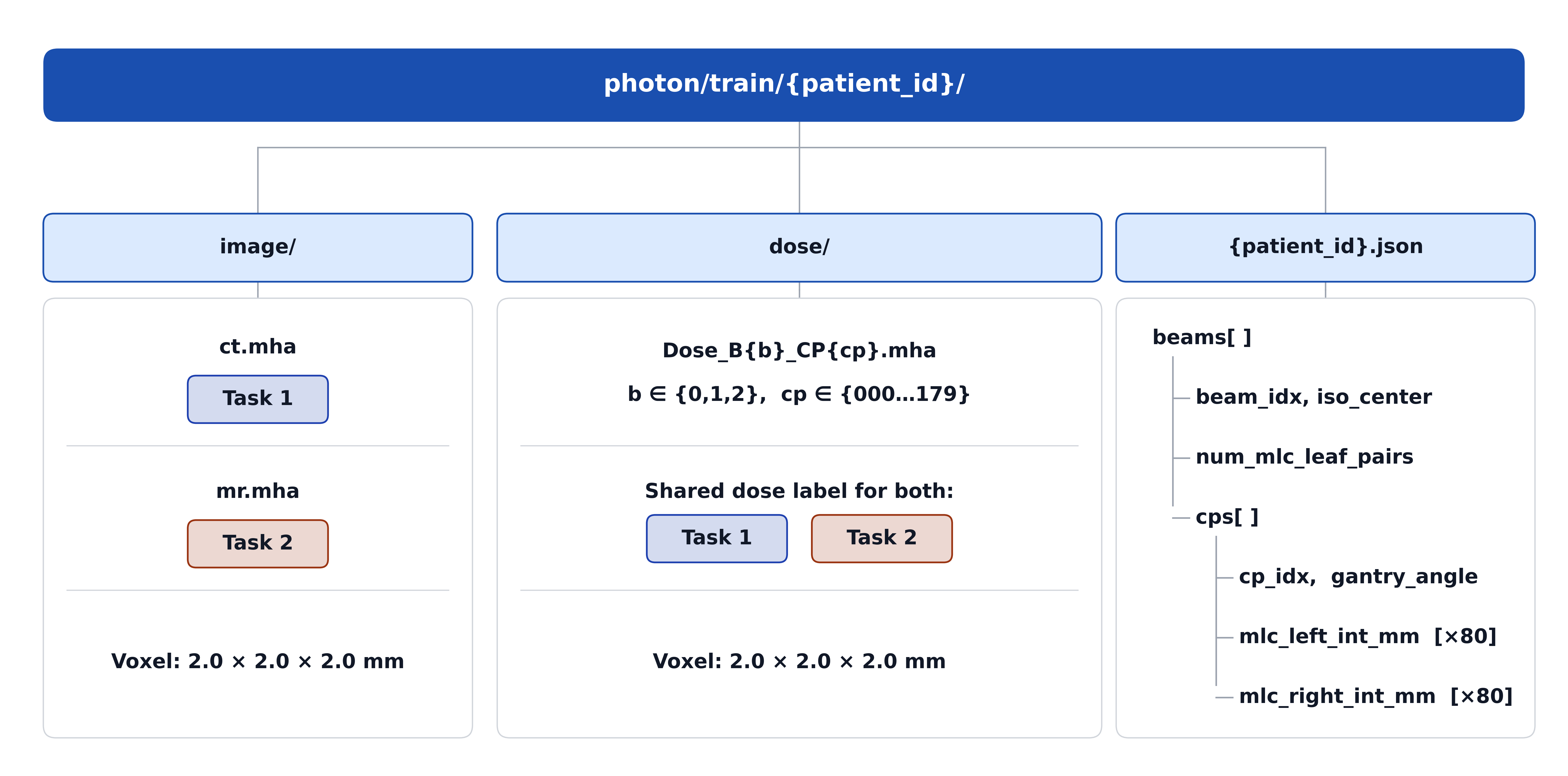}
    \caption{Folder structure for the photon dataset. B, beam; CP, control point.}
    \label{fig:folder_structure_photon}
\end{figure}

\begin{figure}[H]
    \centering
    \includegraphics[width=\textwidth]{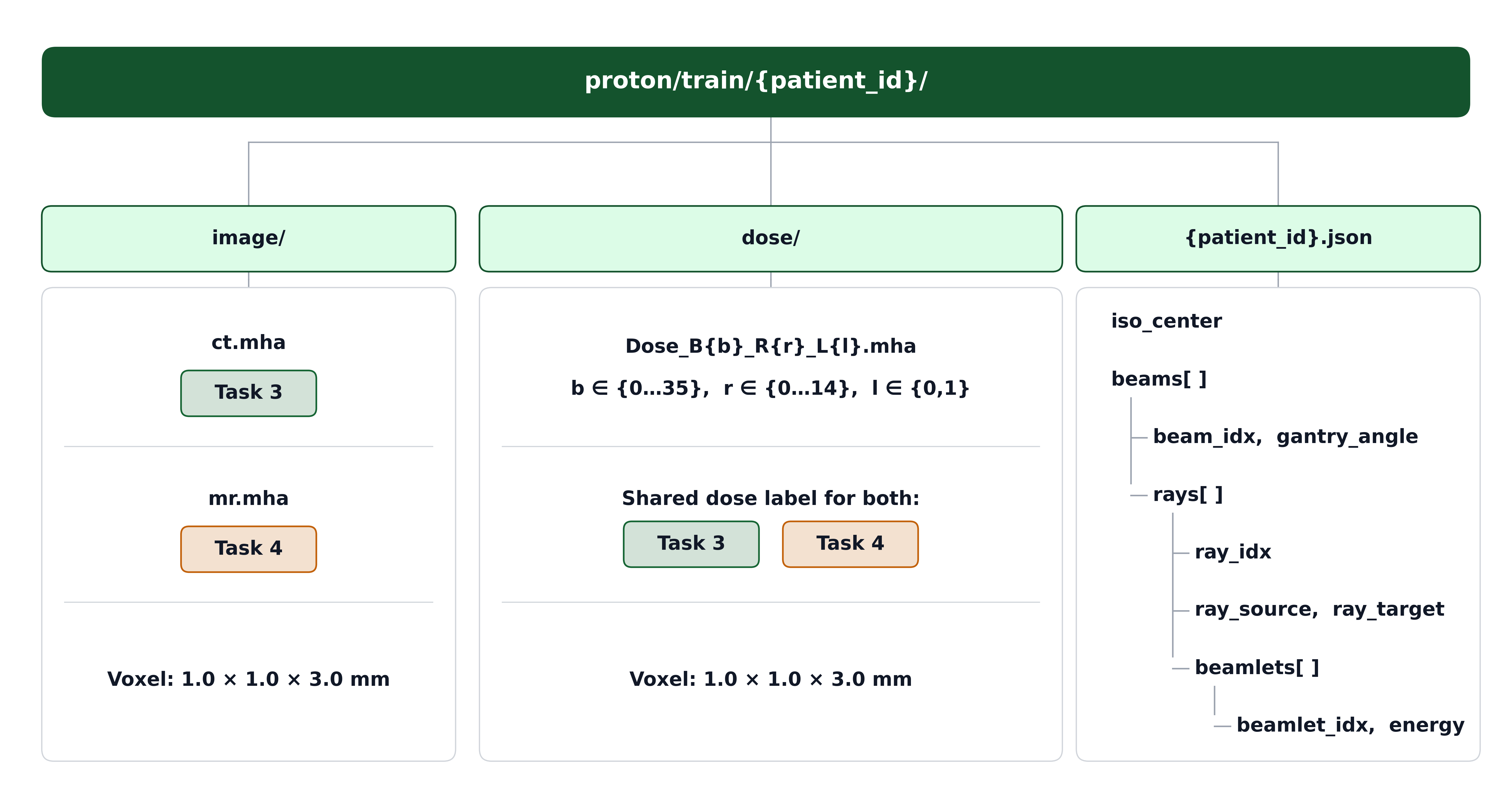}
    \caption{Folder structure for the proton dataset. B, beam; R, ray; L, beamlet.}
    \label{fig:folder_structure_proton}
\end{figure}

The DoseRAD2026 dataset is divided into two subsets, corresponding to photon and proton dose calculation tasks, respectively. Each subset comprises an image directory containing paired \ac{ct}s and \ac{mri}s for each patient, a dose directory with dose maps for multiple beam configurations, and a plan JSON file specifying the associated beam parameters. The corresponding folder structures are illustrated in Figure~\ref{fig:folder_structure_photon} and Figure~\ref{fig:folder_structure_proton}. Each patient is identified by a unique alphanumeric identifier inherited from the SynthRAD2025 dataset.

The dataset is distributed under a CC BY-NC 4.0 International license. The training set was released in April 2026 and the test set will be released in March 2030, after the five-year embargo period of the SynthRAD2025 test data expires. The data are hosted on Hugging Face, with a persistent DOI provided through Zenodo (\url{https://doi.org/10.5281/zenodo.19347848}).

\subsection{Usage notes}

Images and dose maps are provided as compressed MetaImage (.mha) files. These files can be accessed using the Insight Toolkit (ITK) framework or its simplified interfaces, such as SimpleITK, which offer bindings for multiple programming languages, including Python, R, Java, and C++. In addition, MetaImage files can be visualized using commonly available image viewers such as 3D Slicer, vv, and ITK-SNAP.

The JSON file describes the corresponding beam configuration. For photon data, the JSON contains one or more beams (each associated with an isocenter position), with each beam comprising 180 control points defined by gantry angle and 80-pair \ac{mlc} leaf positions (\texttt{mlc\_left\_int\_mm}, \texttt{mlc\_right\_int\_mm}) relative to the beam central axis. For proton data, the JSON specifies the isocenter and 36 beam directions (gantry \ang{0} to \ang{350}, \ang{10} steps), with each beam containing rays defined by source and target coordinates, and the associated beamlet energies in MeV. Dose \texttt{.mha} files follow the naming convention \texttt{Dose\_B\{beam\}\_CP\{cp\}.mha} for photon and \texttt{Dose\_B\{beam\}\_R\{ray\}\_L\{beamlet\}.mha} for proton, directly indexing into the JSON structure. A complete description of all JSON fields is provided in Appendix~\ref{app:json_fields}.

\section{Discussion}

The DoseRAD2026 dataset is the first publicly available dataset providing paired \ac{ct} and \ac{mri} alongside beam-level \ac{mc} dose ground truth for both photon and proton therapy, spanning thoracic and abdominal anatomies across four tasks. It aims to address two clinical needs. First, it provides a large-scale benchmark dataset for \ac{dl}-based acceleration of \ac{mc} dose calculation on \ac{ct}, covering photon segment and proton pencil beam delivery in anatomical sites with challenging tissue heterogeneities. Second, by building on the SynthRAD2025 paired \ac{ct}--\ac{mri} data~\cite{thummerer2025synthrad2025} with additional preprocessing steps including ConvexAdam deformable registration, air cavity correction and systematic manual quality control of \ac{ct}--\ac{mri} alignment, DoseRAD2026 supports exploration of direct \ac{mri} to beam-level dose calculation, offering a potential pathway toward real-time MR-guided or MR-only adaptive radiotherapy. Together, the four tasks provide a common ground for benchmarking \ac{dl} dose calculation methods across modalities, anatomies, and radiation types.

A key strength of the DoseRAD2026 dataset is the provision of beam-level (rather than plan-level) \ac{mc} dose labels, which substantially increases the number of training samples per patient, enables plan dose accumulation, and allows flexible use of patient-specific beam configurations by varying beam angles, isocenters, segment shapes, and proton energies independently. Dose labels for both photon and proton modalities are generated using Geant4, a well-validated general-purpose \ac{mc} simulation framework, providing physically consistent ground truth and a broader scope than existing public datasets such as OpenKBP~\cite{babier2021openkbp}, which covers head-and-neck photon plan doses only. The focus on thoracic and abdominal anatomies introduces clinically relevant tissue heterogeneities, including lung and abdominal air cavities, which present known challenges for analytical dose calculation methods. Full reproducibility is ensured through open-source release of the preprocessing, Geant4 simulation, and matRad \ac{vmat} planning pipelines, together with an analytical pencil beam baseline, allowing the community to scrutinise and reproduce the data generation, or adapt them for their own research purposes.

A fundamental limitation of DoseRAD2026 is that both photon and proton dose distributions are generated using simplified, non-clinical beam models. For photon plans, a generic machine model is used with binary \ac{mlc} aperture masks, without modelling jaw positions, leaf transmission, or tongue-and-groove effects, and dose values are not calibrated to absolute clinical output. The \ac{vmat} plans are generated using matRad \cite{wieser2017development}, an open-source research treatment planning toolkit, which ensures reproducibility but may not capture the full complexity of clinically optimized plans. For proton plans, an idealized generic beam model is employed rather than one specific to any clinical delivery system. Consequently, while the dataset is well-suited for benchmarking \ac{dl} dose calculation algorithms, the dose distributions are not directly representative of clinical deliveries. However, this trade-off is accepted in favor of reproducibility and open-source transparency, the Geant4 \ac{mc} code and matRad planning toolkit ensures that data generation can be verified and extended by the community. Algorithms developed on DoseRAD2026 may nonetheless serve as a strong starting point for clinical translation: the generalized beam representations and physically consistent \ac{mc} ground truth support the learning of fundamental dose–anatomy relationships, and fine-tuning on machine-specific data provides a practical pathway toward clinical deployment. 

Another limitation is that all \ac{mri} data originate from a single center and \ac{mri}-linac system (0.35\,T ViewRay MRidian), which may limit generalisability to other imaging platforms and patient populations. In addition, the dataset relies on deformable registration of \ac{ct} to \ac{mri} to enable supervised learning for direct \ac{mri}-based dose prediction. Despite careful validation and quality control, residual registration errors may persist. Since ground-truth \ac{mc} dose distributions were computed on CT images only and implicitly assumed to be spatially aligned with \ac{mri}, such misregistrations can impact \ac{mri}-based tasks. Residual geometric \ac{mri} distortions following the vendor's corrections were not investigated in this dataset and masking the scans with a union of \ac{ct} and \ac{mri} patient outline masks may further obfuscate their impact on dose calculations. These factors might limit the direct clinical applicability of models trained solely on the provided data and should be considered when interpreting the results or translating methods to clinical workflows.

Beyond the DoseRAD2026 challenge, the dataset will remain publicly available to support continued development and benchmarking of fast CT- and MR-based dose calculation methods. Future extensions may incorporate additional anatomical regions, multi-institutional data, and more clinically realistic beam models to facilitate systematic evaluation of model generalizability and clinical applicability.

\section{Conclusion}
DoseRAD2026 introduces the first publicly available benchmark dataset, providing paired \ac{ct} and \ac{mri} together with beam-level photon and proton \ac{mc} dose distributions for thoracic and abdominal radiotherapy. The dataset and accompanying challenge establish a common framework for the development and rigorous evaluation of fast dose calculation and dose prediction methods across imaging modalities, anatomies, and treatment modalities. We anticipate that the DoseRAD2026 dataset will serve as a valuable open resource for the community and help accelerate progress toward fast, accurate and robust (\ac{dl}-based) dose engines for \ac{ct}- and \ac{mri}-based dose calculation in adaptive radiotherapy.

\section*{Acknowledgements}
The authors would like to thank the GSI Helmholtzzentrum für Schwerionenforschung and the Leibniz-Rechenzentrum for the use of their computing resources. Fan Xiao acknowledges financial support from the China Scholarship Council (No.\ 202308440107). Niklas Wahl and Samir Schulz acknowledge funding from the Deutsche Forschungsgemeinschaft (DFG, German Research Foundation)-Project No. 457509854. Niklas Wahl further acknowledges funding from the Cooperation Program in Cancer Research of the Deutsches Krebsforschungszentrum (DKFZ) and Israel's Ministry of Science, Technology and Space (MOST) as well as from the Deutsche Forschungsgemeinschaft (DFG, German Research Foundation) - Project No. 443188743. The authors would like to thank the Bayerisches Zentrum für Krebsforschung (BZKF) Lighthouse "Image-Guidance in Local Therapies" for supporting the data collection for this dataset.

\section*{Conflict of Interest}
The Department of Radiation Oncology of the University Hospital of LMU Munich has research agreements with Brainlab, Elekta and ViewRay. Zoltan Perko is a Senior Applied Scientist at Radformation Inc.

\section*{References}
\addcontentsline{toc}{section}{\numberline{}References}
\vspace*{-10mm}



\bibliography{./references}      

\bibliographystyle{./medphy.bst}    


\appendix
\section*{Appendices}
\subsection{ConvexAdam registration parameters}
\label{app:convadam_params}

\begin{table}[H]
\centering
\caption{\centering The convexAdam hyperparameters used for MR-to-CT deformable image registration. Detailed description of parameters is provided in Siebert et al.\cite{siebert2025convexadam}.}
\label{tab:convexAdam_params}
\footnotesize
\setlength{\tabcolsep}{6pt}
\begin{tabular}{p{4cm} p{8cm} p{2cm}}
\hline
\textbf{Parameter} & \textbf{Description} & \textbf{Value} \\ \hline
disp\_hwd & The displacement range in voxels of the discretised search space & 6 \\
grid\_sp & The grid spacing in voxels used for the effective resolution of features and deformable transformation & 5 \\
convex\_mind\_r & The MIND radius in voxels & 2 \\
convex\_mind\_d & The MIND dilation in voxels & 2 \\
adam\_mind\_r & The MIND radius in voxels & 1 \\
adam\_mind\_d & The MIND dilation in voxels & 2 \\
adam\_grid\_sp & The grid spacing in voxels for the displacement field resolution & 2 \\
gauss\_\(\sigma\) & $\sigma$ of Gaussian kernel in voxels for B-spline deformation model & 0.2 \\
$\lambda$ & The diffusive regularisation weight & 1 \\
iters & The number of iterations for Adam optimisation & 200 \\
iters\_smooth & The number of post-smoothing (average) steps with a kernel size of 3 & 0 \\
\hline
\end{tabular}
\end{table}

\subsection{Photon and proton beam parameters}
\label{app:beam_params}
\begin{lstlisting}[language=json]
{
  "photon": {
    "SAD_mm": 1000,
    "mlc_num_leaf_pairs": 80,
    "mlc_leaf_thickness_mm": 5,
    "jaw_x_mm": [-200, 200],
    "jaw_y_mm": [-200, 200],
    "source_plane_distance_mm": 2000,
    "virtual_source_distance_mm": 1000,
    "_source_geometry_note": "Photons sampled from a 400x400 mm plane at 2*SAD. Beam converges to a focal point at SAD (1000 mm from isocenter), this focal point is the virtual point source",
    "_mlc_aperture_note": "MLC aperture enforced by rejection sampling against the 400x400 binary mask (1 mm/pixel). History count scales with open area: 10M histories/cm2 of open MLC aperture.",
    "energy_spectrum": {
      "_note": "Geant4 GPS histogram defining photon energy distribution. 100 points, 0.06-6.00 MeV, step 0.06 MeV.",
      "energy_mev": [0.06, 0.12, 0.18, 0.24, 0.30, 0.36, 0.42, 0.48, 0.54, 0.60,
                     0.66, 0.72, 0.78, 0.84, 0.90, 0.96, 1.02, 1.08, 1.14, 1.20,
                     1.26, 1.32, 1.38, 1.44, 1.50, 1.56, 1.62, 1.68, 1.74, 1.80,
                     1.86, 1.92, 1.98, 2.04, 2.10, 2.16, 2.22, 2.28, 2.34, 2.40,
                     2.46, 2.52, 2.58, 2.64, 2.70, 2.76, 2.82, 2.88, 2.94, 3.00,
                     3.06, 3.12, 3.18, 3.24, 3.30, 3.36, 3.42, 3.48, 3.54, 3.60,
                     3.66, 3.72, 3.78, 3.84, 3.90, 3.96, 4.02, 4.08, 4.14, 4.20,
                     4.26, 4.32, 4.38, 4.44, 4.50, 4.56, 4.62, 4.68, 4.74, 4.80,
                     4.86, 4.92, 4.98, 5.04, 5.10, 5.16, 5.22, 5.28, 5.34, 5.40,
                     5.46, 5.52, 5.58, 5.64, 5.70, 5.76, 5.82, 5.88, 5.94, 6.00],
      "weight": [2.54238e-04, 4.19007e-03, 1.64380e-02, 3.27511e-02, 4.26306e-02,
                 4.56607e-02, 4.52659e-02, 4.34762e-02, 4.47129e-02, 3.81616e-02,
                 3.58066e-02, 3.37301e-02, 3.18390e-02, 2.99035e-02, 2.81086e-02,
                 2.64351e-02, 2.49122e-02, 2.34632e-02, 2.21145e-02, 2.09058e-02,
                 1.97536e-02, 1.86973e-02, 1.76865e-02, 1.67660e-02, 1.58873e-02,
                 1.49913e-02, 1.42329e-02, 1.35083e-02, 1.28337e-02, 1.22013e-02,
                 1.16182e-02, 1.10765e-02, 1.05718e-02, 1.00526e-02, 9.61410e-03,
                 9.18486e-03, 8.78546e-03, 8.40149e-03, 8.03662e-03, 7.68496e-03,
                 7.36606e-03, 7.06749e-03, 6.76740e-03, 6.49375e-03, 6.22279e-03,
                 5.97915e-03, 5.75149e-03, 5.51645e-03, 5.29513e-03, 5.07651e-03,
                 4.89121e-03, 4.69359e-03, 4.51351e-03, 4.33559e-03, 4.17315e-03,
                 4.00582e-03, 3.85291e-03, 3.70279e-03, 3.54862e-03, 3.41719e-03,
                 3.27277e-03, 3.15898e-03, 3.02902e-03, 2.90728e-03, 2.79603e-03,
                 2.67682e-03, 2.56908e-03, 2.45770e-03, 2.36121e-03, 2.24374e-03,
                 2.14842e-03, 2.05885e-03, 1.96066e-03, 1.86058e-03, 1.76296e-03,
                 1.68080e-03, 1.59073e-03, 1.50016e-03, 1.41159e-03, 1.32824e-03,
                 1.25038e-03, 1.16362e-03, 1.08056e-03, 1.00508e-03, 9.21253e-04,
                 8.38062e-04, 7.57685e-04, 6.77143e-04, 5.97707e-04, 5.16531e-04,
                 4.31229e-04, 3.53960e-04, 2.77338e-04, 1.94645e-04, 1.14458e-04,
                 2.88556e-05, 5.53396e-08, 7.90565e-09, 0.0, 0.0]
    }
  },

  "proton": {
    "source_model": "point_source",
    "spot_profile": "Gaussian",
    "energy_spread_profile": "Gaussian",
    "_energy_table_note": "85 discrete energies. Each entry: (energy_mev, sigma_energy_mev, sigma_spot_mm). sigma_energy is absolute energy spread (1-sigma). sigma_spot is lateral 1-sigma.",
    "energy_table": [
      {"energy_mev": 31.7290,  "sigma_energy_mev": 6.1130, "sigma_spot_mm": 7.4934},
      {"energy_mev": 36.7986,  "sigma_energy_mev": 5.1891, "sigma_spot_mm": 7.6244},
      {"energy_mev": 41.3788,  "sigma_energy_mev": 4.5578, "sigma_spot_mm": 7.7018},
      {"energy_mev": 45.5978,  "sigma_energy_mev": 4.0937, "sigma_spot_mm": 7.7653},
      {"energy_mev": 49.5354,  "sigma_energy_mev": 3.7352, "sigma_spot_mm": 7.8180},
      {"energy_mev": 53.2453,  "sigma_energy_mev": 3.4482, "sigma_spot_mm": 7.8620},
      {"energy_mev": 56.7659,  "sigma_energy_mev": 3.2122, "sigma_spot_mm": 7.8988},
      {"energy_mev": 60.1259,  "sigma_energy_mev": 3.0139, "sigma_spot_mm": 7.9294},
      {"energy_mev": 63.3471,  "sigma_energy_mev": 2.8446, "sigma_spot_mm": 7.9546},
      {"energy_mev": 66.4468,  "sigma_energy_mev": 2.6979, "sigma_spot_mm": 7.9752},
      {"energy_mev": 69.4389,  "sigma_energy_mev": 2.5693, "sigma_spot_mm": 7.9916},
      {"energy_mev": 72.3349,  "sigma_energy_mev": 2.4554, "sigma_spot_mm": 7.9938},
      {"energy_mev": 75.1442,  "sigma_energy_mev": 2.3537, "sigma_spot_mm": 7.7121},
      {"energy_mev": 77.8749,  "sigma_energy_mev": 2.2623, "sigma_spot_mm": 7.5646},
      {"energy_mev": 80.5337,  "sigma_energy_mev": 2.1795, "sigma_spot_mm": 7.4232},
      {"energy_mev": 83.1266,  "sigma_energy_mev": 2.1041, "sigma_spot_mm": 7.2877},
      {"energy_mev": 85.6587,  "sigma_energy_mev": 2.0351, "sigma_spot_mm": 7.1577},
      {"energy_mev": 88.1344,  "sigma_energy_mev": 1.9716, "sigma_spot_mm": 7.0329},
      {"energy_mev": 90.5576,  "sigma_energy_mev": 1.9130, "sigma_spot_mm": 6.9130},
      {"energy_mev": 92.9319,  "sigma_energy_mev": 1.8586, "sigma_spot_mm": 6.7977},
      {"energy_mev": 95.2605,  "sigma_energy_mev": 1.8081, "sigma_spot_mm": 6.6868},
      {"energy_mev": 97.5459,  "sigma_energy_mev": 1.7610, "sigma_spot_mm": 6.5796},
      {"energy_mev": 99.7909,  "sigma_energy_mev": 1.7169, "sigma_spot_mm": 6.4764},
      {"energy_mev": 101.9976, "sigma_energy_mev": 1.6756, "sigma_spot_mm": 6.3771},
      {"energy_mev": 104.1682, "sigma_energy_mev": 1.6367, "sigma_spot_mm": 6.2814},
      {"energy_mev": 106.3045, "sigma_energy_mev": 1.6000, "sigma_spot_mm": 6.1893},
      {"energy_mev": 108.4082, "sigma_energy_mev": 1.5654, "sigma_spot_mm": 6.1006},
      {"energy_mev": 110.4810, "sigma_energy_mev": 1.5326, "sigma_spot_mm": 6.0151},
      {"energy_mev": 112.5242, "sigma_energy_mev": 1.5015, "sigma_spot_mm": 5.9327},
      {"energy_mev": 114.5392, "sigma_energy_mev": 1.4721, "sigma_spot_mm": 5.8533},
      {"energy_mev": 116.5273, "sigma_energy_mev": 1.4440, "sigma_spot_mm": 5.7768},
      {"energy_mev": 118.4897, "sigma_energy_mev": 1.4173, "sigma_spot_mm": 5.7029},
      {"energy_mev": 120.4273, "sigma_energy_mev": 1.3918, "sigma_spot_mm": 5.6316},
      {"energy_mev": 122.3412, "sigma_energy_mev": 1.3674, "sigma_spot_mm": 5.5628},
      {"energy_mev": 124.2323, "sigma_energy_mev": 1.3441, "sigma_spot_mm": 5.4966},
      {"energy_mev": 126.1016, "sigma_energy_mev": 1.3218, "sigma_spot_mm": 5.4328},
      {"energy_mev": 127.9497, "sigma_energy_mev": 1.3004, "sigma_spot_mm": 5.3712},
      {"energy_mev": 129.7775, "sigma_energy_mev": 1.2799, "sigma_spot_mm": 5.3119},
      {"energy_mev": 131.5856, "sigma_energy_mev": 1.2602, "sigma_spot_mm": 5.2547},
      {"energy_mev": 133.3749, "sigma_energy_mev": 1.2412, "sigma_spot_mm": 5.1996},
      {"energy_mev": 135.1458, "sigma_energy_mev": 1.2229, "sigma_spot_mm": 5.1465},
      {"energy_mev": 136.8991, "sigma_energy_mev": 1.2054, "sigma_spot_mm": 5.0953},
      {"energy_mev": 138.6352, "sigma_energy_mev": 1.1884, "sigma_spot_mm": 5.0459},
      {"energy_mev": 140.3548, "sigma_energy_mev": 1.1720, "sigma_spot_mm": 4.9983},
      {"energy_mev": 142.0583, "sigma_energy_mev": 1.1562, "sigma_spot_mm": 4.9525},
      {"energy_mev": 143.7462, "sigma_energy_mev": 1.1409, "sigma_spot_mm": 4.9083},
      {"energy_mev": 145.4189, "sigma_energy_mev": 1.1261, "sigma_spot_mm": 4.8657},
      {"energy_mev": 147.0770, "sigma_energy_mev": 1.1118, "sigma_spot_mm": 4.8247},
      {"energy_mev": 148.7208, "sigma_energy_mev": 1.0980, "sigma_spot_mm": 4.7852},
      {"energy_mev": 150.3508, "sigma_energy_mev": 1.0845, "sigma_spot_mm": 4.7472},
      {"energy_mev": 151.9672, "sigma_energy_mev": 1.0715, "sigma_spot_mm": 4.7105},
      {"energy_mev": 153.5705, "sigma_energy_mev": 1.0589, "sigma_spot_mm": 4.6752},
      {"energy_mev": 155.1611, "sigma_energy_mev": 1.0466, "sigma_spot_mm": 4.6412},
      {"energy_mev": 156.7391, "sigma_energy_mev": 1.0348, "sigma_spot_mm": 4.6084},
      {"energy_mev": 158.3051, "sigma_energy_mev": 1.0231, "sigma_spot_mm": 4.5768},
      {"energy_mev": 159.8591, "sigma_energy_mev": 1.0119, "sigma_spot_mm": 4.5464},
      {"energy_mev": 161.4017, "sigma_energy_mev": 1.0010, "sigma_spot_mm": 4.5172},
      {"energy_mev": 162.9330, "sigma_energy_mev": 0.9903, "sigma_spot_mm": 4.4889},
      {"energy_mev": 164.4532, "sigma_energy_mev": 0.9800, "sigma_spot_mm": 4.4618},
      {"energy_mev": 165.9628, "sigma_energy_mev": 0.9699, "sigma_spot_mm": 4.4356},
      {"energy_mev": 167.4618, "sigma_energy_mev": 0.9601, "sigma_spot_mm": 4.4104},
      {"energy_mev": 168.9506, "sigma_energy_mev": 0.9503, "sigma_spot_mm": 4.3860},
      {"energy_mev": 170.4293, "sigma_energy_mev": 0.9409, "sigma_spot_mm": 4.3626},
      {"energy_mev": 171.8982, "sigma_energy_mev": 0.9319, "sigma_spot_mm": 4.3399},
      {"energy_mev": 173.3576, "sigma_energy_mev": 0.9230, "sigma_spot_mm": 4.3181},
      {"energy_mev": 174.8075, "sigma_energy_mev": 0.9142, "sigma_spot_mm": 4.2970},
      {"energy_mev": 176.2482, "sigma_energy_mev": 0.9057, "sigma_spot_mm": 4.2767},
      {"energy_mev": 177.6799, "sigma_energy_mev": 0.8975, "sigma_spot_mm": 4.2571},
      {"energy_mev": 179.1028, "sigma_energy_mev": 0.8892, "sigma_spot_mm": 4.2381},
      {"energy_mev": 180.5170, "sigma_energy_mev": 0.8815, "sigma_spot_mm": 4.2198},
      {"energy_mev": 181.9227, "sigma_energy_mev": 0.8736, "sigma_spot_mm": 4.2021},
      {"energy_mev": 183.3202, "sigma_energy_mev": 0.8660, "sigma_spot_mm": 4.1849},
      {"energy_mev": 184.7095, "sigma_energy_mev": 0.8585, "sigma_spot_mm": 4.1682},
      {"energy_mev": 186.0907, "sigma_energy_mev": 0.8514, "sigma_spot_mm": 4.1521},
      {"energy_mev": 187.4642, "sigma_energy_mev": 0.8442, "sigma_spot_mm": 4.1364},
      {"energy_mev": 188.8299, "sigma_energy_mev": 0.8373, "sigma_spot_mm": 4.1212},
      {"energy_mev": 190.1880, "sigma_energy_mev": 0.8304, "sigma_spot_mm": 4.1063},
      {"energy_mev": 191.5388, "sigma_energy_mev": 0.8236, "sigma_spot_mm": 4.0919},
      {"energy_mev": 192.8822, "sigma_energy_mev": 0.8170, "sigma_spot_mm": 4.0779},
      {"energy_mev": 194.2185, "sigma_energy_mev": 0.8107, "sigma_spot_mm": 4.0641},
      {"energy_mev": 195.5477, "sigma_energy_mev": 0.8043, "sigma_spot_mm": 4.0507},
      {"energy_mev": 196.8700, "sigma_energy_mev": 0.7981, "sigma_spot_mm": 4.0375},
      {"energy_mev": 198.1855, "sigma_energy_mev": 0.7919, "sigma_spot_mm": 4.0246},
      {"energy_mev": 199.4943, "sigma_energy_mev": 0.7860, "sigma_spot_mm": 4.0119},
      {"energy_mev": 200.7966, "sigma_energy_mev": 0.7803, "sigma_spot_mm": 3.9994}
    ]
  },

  "hu_to_density": {
    "_note": "HU-to-density calibration curve used by G4DCM. 10 anchor points linearly interpolated. Both photon and proton simulations use identical tables.",
    "entries": [
      {"hu": -1024, "density_g_cm3": 1.200000e-03},
      {"hu":  -999, "density_g_cm3": 1.210000e-03},
      {"hu":  -200, "density_g_cm3": 8.043754e-01},
      {"hu":  -199, "density_g_cm3": 8.183035e-01},
      {"hu":   -10, "density_g_cm3": 1.006579e+00},
      {"hu":    -9, "density_g_cm3": 9.966749e-01},
      {"hu":   120, "density_g_cm3": 1.126553e+00},
      {"hu":   121, "density_g_cm3": 1.095097e+00},
      {"hu":  3000, "density_g_cm3": 3.027294e+00},
      {"hu":  4000, "density_g_cm3": 3.698428e+00}
    ]
  }
}
\end{lstlisting}

\subsection{Patient IDs with metal artifacts}
\label{app:metal_ids}

Metal artifacts were observed in the following training cases:
\begin{itemize}
\setlength{\itemsep}{0pt}
\setlength{\topsep}{0pt}
\item \texttt{1THB016}
\item \texttt{1THB021}
\item \texttt{1THB029}
\item \texttt{1THB031}
\item \texttt{1THB052}
\item \texttt{1THB054}
\item \texttt{1THB067}
\item \texttt{1THB078}
\item \texttt{1THB122}
\item \texttt{1ABB138}
\item \texttt{1THB214}
\item \texttt{1THB217}
\item \texttt{1ABB078}
\end{itemize}

\subsection{Plan JSON field descriptions}
\label{app:json_fields}

\vspace{1em}
\begin{table}[H]
\centering
\caption{Photon plan JSON field description.}
\label{tab:json_photon}
\footnotesize
\setlength{\tabcolsep}{6pt}
\begin{tabular}{p{4.0cm} p{1.8cm} p{8.2cm}}
\hline
\textbf{Field} & \textbf{Unit} & \textbf{Description} \\ \hline
\multicolumn{3}{l}{\textit{Per beam}} \\
\texttt{beam\_idx} & -- & Sequential beam index (0-indexed) \\
\texttt{iso\_center} & mm & 3D coordinate of the beam rotation center in patient \\
\texttt{num\_mlc\_leaf\_pairs} & -- & Number of leaf pairs to shape the segment \\
\texttt{control\_points} & -- & Discrete beam delivery sampling points along the arc \\ \hline
\multicolumn{3}{l}{\textit{Per control point}} \\
\texttt{cp\_idx} & -- & Control point index within the arc (0-indexed) \\
\texttt{gantry\_angle} & deg & Angle of the photon beam direction around patient \\
\texttt{mlc\_left\_int\_mm} & mm & Left-edge positions of 80 collimator leaves\\
\texttt{mlc\_right\_int\_mm} & mm & Right-edge positions of 80 collimator leaves\\ \hline
\end{tabular}
\end{table}

\vspace{3em}
\begin{table}[H]
\centering
\caption{Proton plan JSON field description.}
\label{tab:json_proton}
\footnotesize
\setlength{\tabcolsep}{6pt}
\begin{tabular}{p{4.0cm} p{1.8cm} p{8.2cm}}
\hline
\textbf{Field} & \textbf{Unit} & \textbf{Description} \\ \hline
\texttt{iso\_center} & mm & 3D coordinate of the beam rotation center in patient \\ \hline
\multicolumn{3}{l}{\textit{Per beam}} \\
\texttt{beam\_idx} & -- & Sequential beam index (0-indexed) \\
\texttt{gantry\_angle} & deg & Angle of the proton beam direction around patient \\
\texttt{rays} & -- & A group of pencil beamlets from the same BEV grid position within the beam\\ \hline
\multicolumn{3}{l}{\textit{Per ray}} \\
\texttt{ray\_idx} & -- & Ray index within the beam (0-indexed) \\
\texttt{ray\_source} & mm & Source point of the ray on the source plane\\
\texttt{ray\_target} & mm & Aim point of the ray on the isocenter plane\\
\texttt{beamlets} & -- & Energy layers for the ray \\ \hline
\multicolumn{3}{l}{\textit{Per beamlet}} \\
\texttt{beamlet\_idx} & -- & Energy layer index (0-indexed) \\
\texttt{energy} & MeV & Proton energy, determining the penetration depth \\ \hline
\end{tabular}
\end{table}

\end{document}

%% file: medphy.tex
\usepackage[super,sort&compress,comma]{natbib}

\usepackage{fancyhdr}		

\usepackage[section]{placeins}   %

\usepackage{graphicx}

\makeatletter \renewcommand\@biblabel[1]{$^{#1}$} \makeatother
\newcommand{\cen}[1]{\begin{center} #1 \end{center}}

\usepackage[mathlines]{lineno}

\newcommand*\patchAmsMathEnvironmentForLineno[1]{%
	\expandafter\let\csname old#1\expandafter\endcsname\csname #1\endcsname
	\expandafter\let\csname oldend#1\expandafter\endcsname\csname end#1\endcsname
	\renewenvironment{#1}%
	{\linenomath\csname old#1\endcsname}%
	{\csname oldend#1\endcsname\endlinenomath}}%
\newcommand*\patchBothAmsMathEnvironmentsForLineno[1]{%
	\patchAmsMathEnvironmentForLineno{#1}%
	\patchAmsMathEnvironmentForLineno{#1*}}%
\AtBeginDocument{%
	\patchBothAmsMathEnvironmentsForLineno{equation}%
	\patchBothAmsMathEnvironmentsForLineno{align}%
	\patchBothAmsMathEnvironmentsForLineno{flalign}%
	\patchBothAmsMathEnvironmentsForLineno{alignat}%
	\patchBothAmsMathEnvironmentsForLineno{gather}%
	\patchBothAmsMathEnvironmentsForLineno{multline}%
}

\usepackage{hyperref}
\hypersetup{ colorlinks,
    citecolor=blue,
    filecolor=blue,
    linkcolor=blue,
    urlcolor=blue
}

\usepackage{xcolor}

\definecolor{gray}{rgb}{0.6,0.6,0.6}
\definecolor{red}{rgb}{0.85,0,0}
\definecolor{green}{rgb}{0,0.85,0}
\definecolor{blue}{rgb}{0,0,0.85}
\definecolor{beige}{rgb}{0.92,0.87,0.78}
\usepackage[all]{hypcap}    